# Ferromagnetic order at room temperature in monolayer WSe$_2$ semiconductor via vanadium dopant

Seok Joon Yun[1†*], Dinh Loc Duong[1†*], Doan Manh Ha[1,2], Kirandeep Singh[1], Thanh Luan Phan[1,3], Wooseon Choi[2], Young-Min Kim[1,2], and Young Hee Lee[1,2,4*]

[1]Center for Integrated Nanostructure Physics (CINAP), Institute for Basic Science (IBS), Suwon 16419, Republic of Korea.

[2]Department of Energy Science, Sungkyunkwan University, Suwon 16419, Republic of Korea.

[3]Department of Electronic and Electrical Engineering, Sungkyunkwan University, Suwon 16419, Republic of Korea

[4]Department of Physics, Sungkyunkwan University, Suwon 16419, Republic of Korea.

[†]These authors contributed equally to this work.

[*]email: tjrwns07@skku.edu, ddloc@skku.edu, leeyoung@skku.edu

**Diluted magnetic semiconductors including Mn-doped GaAs are attractive for gate-controlled spintronics but Curie transition at room temperature with long-range ferromagnetic order is still debatable to date**[1–7]**. Here, we report the room-temperature ferromagnetic domains with long-range order in semiconducting V-doped WSe$_2$ monolayer synthesized by chemical vapor deposition. Ferromagnetic order is manifested using magnetic force microscopy up to 360K, while retaining high on/off current ratio of ~10$^5$ at 0.1% V-doping concentration. The V-substitution to W sites keep a V-V separation distance of 5 nm without V-V aggregation, scrutinized by high-resolution scanning transmission-electron-microscopy, which implies the possibility of the Ruderman-Kittel-Kasuya-Yoshida interaction (or Zener model) by establishing the long-range ferromagnetic order in V-doped WSe$_2$ monolayer through free hole carriers. More importantly, the ferromagnetic order is clearly modulated by applying a back gate. Our findings open new opportunities for using two-dimensional transition metal dichalcogenides for future spintronics.**

The magnetism of semiconductors can be tuned by modulating carrier concentration with a gate bias. While intrinsic magnetic semiconductors rarely exist in nature[8], incorporation of magnetic dopants into innumerable semiconductors, called diluted magnetic semiconductors (DMSs), allows us to construct an inherent ferromagnetic (FM) state with spin-polarized carriers at the Fermi level[1–7]. The most typical example is Mn-doped GaAs, which exhibits a gate-controlled magnetic hysteresis, yielding a large number of spintronic devices such as spin-injection sources and memory devices[1,9–13]. Nevertheless, the Curie temperature ($T_c$) of ferromagnetic transition in magnetic semiconductors is scarcely accessible to room-temperature (RT), precluding the use of these materials to practical implementations[1–4,14]. The ferromagnetic state in magnetic metal-doped oxides and nitrides is available at RT but is localized to aggregated metal oxide/nitride nanoparticles without a long-range magnetic order[3].

The ferromagnetic state in van der Waals two-dimensional (2D) materials has been observed recently in the monolayer limit[15–21]. Intrinsic $CrI_3$ and $CrGeTe_3$ semiconductors reveal ferromagnetism but the $T_c$ is still low below 60K[20,21]. In contrast, monolayer $VSe_2$ and $MnSe_2$ are ferromagnetic metals with $T_c$ above RT but incapable of controlling its carrier density[22,23]. Moreover, the long-range ferromagnetic order in doped diluted chalcogenide semiconductors has not been demonstrated at RT[24–28]. The key research target is to realize the long-range order ferromagnetism, $T_c$ over RT, and semiconductor with gate tunability. Here, we unambiguously observe tunable magnetic domains by a gate bias above RT in diluted V-doped $WSe_2$, while maintaining the semiconducting characteristic of $WSe_2$ with a high on/off current ratio of five orders of magnitude.

Figure 1a illustrates the schematic for the synthesis of V-doped monolayer $WSe_2$ via chemical vapor deposition (CVD). A metal precursor solution prepared by mixing V and W liquid sources at a given atomic ratio was spin-coated on $SiO_2$ substrate and the substrate was introduced into the CVD chamber with selenium. The metal precursors get decomposed into metal oxides at growth temperature, resulting in monolayer $V_xW_{1-x}Se_2$, followed by selenization. The atomic ratio of V to W sources in precursor solution can be precisely controlled from 0.1% to 40%, while the hexagonal flakes are retained in a monolayer form (the optical image in Fig. 1a and Supplementary Fig. S1). Meanwhile, the dendritic and multilayer flakes are partially generated at higher V-concentration. The V atoms are incorporated into monolayer $WSe_2$ with V/W contents similar to nominal values, as confirmed by X-ray photoelectron spectroscopy (Supplementary Fig. S2).

To study the doping effect of vanadium to the electronic properties of WSe$_2$, field effect transistors (FETs) of V-doped monolayer WSe$_2$ were fabricated (Fig. 1b). The CVD-grown pristine WSe$_2$ manifests a *p*-type semiconductor with a threshold voltage at −50 V. The threshold voltage is shifted to −10 V for 0.5% V-doped sample and further increased to +20 V for 1% V-doped sample, while retaining high on/off ratio of ~$10^5$ and distinct *p*-type behavior. In contrast, the on/off ratio is significantly reduced for >10% doping samples due to the formation of heavily degenerate V-doped WSe$_2$.

Since V-doped WSe$_2$ monolayer sample has a extremely small mass ($10^{-6}$ gram per cm$^2$), mass-dependent magnetic characterization methods such as vibrating sample magnetometer (VSM) and superconducting quantum interference device (SQUID) are not adequate due to the detection limit. The signal from V-doped WSe$_2$ monolayer sample could be below the detection limit of these systems (Supplementary Fig. S3). Furthermore, such small signals, if detectable, may be obscured by the artifacts of magnetic impurities (e.g. particles with iron content, kapton tape for holding samples) during sample preparation and measurement processes[29].

Meanwhile, the magnetic force microscopy (MFM) is surface-sensitive and can detect very small magnetic dipole moments with a nanometer scale (Supplementary Fig. S3). Therefore, we investigate the existence of magnetic order of V-doped WSe$_2$ monolayer by using the MFM. Figures 1c-f are the tapping-mode topography and MFM phase images with a Co-Cr tip of pristine WSe$_2$ (Fig. 1c-d) and 0.1% V-doped (Fig. 1e-f) WSe$_2$ monolayer at 300K. No distinctive feature was observed in MFM phase, indicating no ferromagnetic order in pristine WSe$_2$ (Fig. 1d). In contrast, three distinct features (marked as numbers) are eminent in MFM phase image at 0.1% V-doped WSe$_2$ (Fig. 1f and the schematic in Fig. 1g) : (i) large domains with phase contrast separated by domain walls (white dotted line), (ii) the dendritic patterns in monolayer flake and (iii) multilayer dendrites, correlated with the topography image in Figure 1e.

Ferromagnetic domain stripes of the MFM phase are more clearly manifested at 150K (Fig. 2a). The domain stripes merge (region a) and split (region b) as temperature increases, strongly implying the domain features originated from magnetic response (Fig. 2b and Supplementary Fig. S4). The distinct magnetic phase becomes ambiguous above RT (Fig. 2c) but the magnetic domain walls still retain reminent up to 420K. To elucidate how the magnetic domains are modulated with V-doping concentration, we further conducted MFM for 0.5% and 2% V-doped $WSe_2$ (Fig. 2d and Supplementary Fig. S5). A strong contrast phase within the flake still emerges from the 0.5% sample up to 360K with dendritic patterns. Similar dendritic patterns were observed in the photoluminescence (PL) mapping (Fig. 2e). The PL intensity quenching and peak red-shift in region (2) and (3) with respect to region (1) originate from the formation of positive trions and charge screening (Fig. 2f). This is ascribed to inhomogeneous V-doping concentration in different regions. This does not necessarily imply strong correlation of the magnetic domains to the PL patterns. The PL measurements are also consistent to Raman mapping (Supplementary Fig. S6, S7).

To further confirm the magnetic nature of the observed domains, we carried out the MFM measurements with the tip magnetized vertically and horizontally. The amplitude of the magnetic signals is much stronger with vertical magnetized tip than with horizontal one (Fig. 2g, h). Additionally, the sign of the phase (compared to the $SiO_2$ background) is inverted in some domains. For example, the contrast of region (b) is positive with vertical magnetized tip, whereas it is negative with horizontal one. This indicates the magnetic force changes from repulsive to attractive force, which is also solid evidence for the magnetic response. Tip-dependent MFM measurement was further performed to ensure magnetism of the V-doped WSe$_2$ (Supplementary Fig. S8).

Figure 2i summarizes the magnetic response of V-doped WSe$_2$ samples with different V-doping concentrations. At low doping concentration (0.1% and 0.3%), the magnetic domains were retained with a line stripe. The magnetic domains were transformed to a polygon shape at high V-doping concentration (0.5% and 2%). We note that that the area of the magnetic domain becomes smaller as the V-doping concentration increases and the phase contrast was eventually not appreciable at 10%. The magnetic domains are also sensitive under ambient conditions (Supplementary Fig. S9).

To investigate the atomic structure of V-doped WSe$_2$, we conducted the annular-dark-field scanning–transmission-electron-microscopy (ADF-STEM) for 0.1% and 2% V-doped WSe$_2$ (Fig. 3). A well-crystallized 2H–WSe$_2$ structure is clearly demonstrated with W (bright) and Se (grey) sites with additional V (dark) atoms, as is consistent with the intensity profile with simulated images (Fig. 3a-d). Well-distributed V atoms in both samples without any clustering indicates that the magnetic properties of V-doped WSe$_2$ did not originate from phase segregation of vanadium but rather result from interaction between V atoms via host WSe$_2$. We analyzed the concentration for 0.1% and 2% V-doped samples (Fig. 3e, f). Three impurities are clearly identified after Wiener-filter false coloring (Supplementary Fig. S10): Se-vacancy in WSe$_2$ (WSe), V-substitution into W site in WSe (VSe), and VSe$_2$. Five STEM images of each sample were thoughtfully analyzed for reliable statistics (Supplementary Fig. S11). The real V-concentration of nominal 0.1% and 2% V-doped samples are similar to $0.1 \pm 0.01\%$ and $1.0 \pm 0.07\%$, respectively (Supplementary Fig. S12).

Chalcogen defects can influence magnetic properties of TMDs[31]. However, Se-vacancy concentration is irrespective of the V composition (approximate 1% or $10^{13}$ cm$^{-2}$ in both 0.1% and 2% samples). This strongly implies that the RT-ferromagnetism is not attributed by Se-vacancies. The role of V-substituted forms (VSe or VSe$_2$) is as yet unclear for the magnetism of V-doped WSe$_2$. However, the ratios of VSe to total V atoms (VSe+VSe$_2$) are similar in both 0.1% and 2% V-doped samples (Supplementary Fig. S13), indicating that VSe or VSe$_2$ doping concentration is closely correlated to each other for discernible $T_c$. We next explore the first and second nearest neighbor distances between V atoms (Fig. 3h). The average V-V neighbor distance is much longer in 0.1% (50 Å, or ~ 15 unit cells) than in 2% (18 Å, or ~5 unit cells) V-doped samples.

Gate-tunable magnetic property is crucial evidence to demonstrate the ferromagnetic semiconductor. We performed the MFM measurements under applying gate biases (Fig. 4a). The phase contrasts by ferromagnetic domains of 0.1% V-doped WSe$_2$ with different gate biases from -10 V to 20 V are shown in Fig. 4b. This demonstrates an apparent variation of the contrast between domains with gate biases (Supplementary Fig. S14). The deviation of contrast between domains is almost negligible at a negative gate bias of -10 V. As the gate bias is shifted toward the positive gate bias to 15 V, the phase contrast becomes distinct (Fig. 4c). Interestingly, the phase deviation drops at a high gate bias of > 15 V. The non-monotonic change of the phase deviation through the gate bias again excludes the electrostatic artifact in our gate-dependent MFM measurement (if the phase deviation is solely contributed by electrostatic force, it should be monotonic with the gate bias). Furthermore, the faint magnetic domains at the negative bias (-10 V) at room temperature emerged predominantly at 150K (Fig. 4d-e). This strongly indicates the gate-tunable magnetic properties in V-doped WSe$_2$ samples.

In summary, we have successfully synthesized V-doped WSe$_2$ in monolayer, which reveals RT dilute ferromagnetic semiconductors. The existence of the ferromagnetic order is confirmed at microscopic scale by MFM. Furthermore, the magnetic order can be modulated with back-gate bias, indicating the validation of the RKKY model in establishing the long-range magnetic order. Our work opens a direct route to demonstrate practical applications of TMDs in spintronic devices at RT.

## Acknowledgements

This work was supported by the Institute for Basic Science (IBS-R011-D1). We are grateful for useful discussion with Sung Wng Kim at CINAP, SKKU, Manh-Huong Phan at USF and Philip Kim at Harvard. The computational supports are provided from the Korea Institute of Science and Technology Information (KISTI) under grants numbers KSC-2016-C3-0042 and KSC-2017-C2-0056.

## Author contributions

S.J.Y., and D.L.D. initiated this work. D.M.H. fabricated and characterized FET devices. K.S. analyzed VSM data. L.P. performed Al$_2$O$_3$-passivation. W.S.C., and Y.-M.K. performed TEM measurement and analysis. Y.H.L. guided and analyzed the work. S.J.Y., D.L.D. and Y.H.L. wrote the manuscript. All authors participated in the manuscript review.

## Competing interests

The authors declare no competing interests.

## Additional information

**Supplementary information** is available for this paper at http://doi.org/xxx.

**Correspondence and requests for materials** should be addressed to S.J.Y. (tjrwns07@skku.edu), D.L.D. (ddloc@skku.edu), Y.H.L. (leeyoung@skku.edu).

# References

1. Dietl, T. & Ohno, H. Dilute ferromagnetic semiconductors: Physics and spintronic structures. *Rev. Mod. Phys.* **86,** 187–251 (2014).

2. Awschalom, D. D. & Flatté, M. E. Challenges for semiconductor spintronics. *Nat. Phys.* **3,** 153–159 (2007).

3. Dietl, T. A ten-year perspective on dilute magnetic semiconductors and oxides. *Nat. Mater.* **9,** 965–974 (2010).

4. Macdonald, A. H., Schiffer, P. & Samarth, N. Ferromagnetic semiconductors: Moving beyond (Ga,Mn)As. *Nat. Mater.* **4,** 195–202 (2005).

5. Jungwirth, T., Sinova, J., Mašek, J., Kučera, J. & MacDonald, A. H. Theory of ferromagnetic (III,Mn)V semiconductors. *Rev. Mod. Phys.* **78,** 809–864 (2006).

6. Sato, K. *et al.* First-principles theory of dilute magnetic semiconductors. *Rev. Mod. Phys.* **82,** 1633–1690 (2010).

7. Bonanni, A. & Dietl, T. A story of high-temperature ferromagnetism in semiconductors. *Chem. Soc. Rev.* **39,** 528–539 (2010).

8. Dietl, T. Ferromagnetic semiconductors. *Semicond. Sci. Technol.* **17,** 377–392 (2002).

9. Pappert, K. *et al.* Magnetization-Switched Metal-Insulator Transition in a (Ga,Mn)As Tunnel Device. *Phys. Rev. Lett.* **97,** 186402 (2006).

10. Mark, S. *et al.* Fully Electrical Read-Write Device Out of a Ferromagnetic Semiconductor.

*Phys. Rev. Lett.* **106,** 057204 (2011).

11. Kohda, M., Kita, T., Ohno, Y., Matsukura, F. & Ohno, H. Bias voltage dependence of the electron spin injection studied in a three-terminal device based on a (Ga,Mn)As∕n+-GaAs Esaki diode. *Appl. Phys. Lett.* **89,** 012103 (2006).

12. Tanaka, M. & Higo, Y. Large Tunneling Magnetoresistance in GaMnAs/AlAs/GaMnAs Ferromagnetic Semiconductor Tunnel Junctions. *Phys. Rev. Lett.* **87,** 026602 (2001).

13. Ohno, H. *et al.* Electric-field control of ferromagnetism. *Nature* **408,** 944–946 (2000).

14. Dietl, T. Zener Model Description of Ferromagnetism in Zinc-Blende Magnetic Semiconductors. *Science (80-. ).* **287,** 1019–1022 (2000).

15. Bonilla, M. *et al.* Strong room-temperature ferromagnetism in VSe2 monolayers on van der Waals substrates. *Nat. Nanotechnol.* **13,** 289–294 (2018).

16. Klein, D. R. *et al.* Probing magnetism in 2D van der Waals crystalline insulators via electron tunneling. *Science (80-. ).* **360,** 1218–1222 (2018).

17. Song, T. *et al.* Giant tunneling magnetoresistance in spin-filter van der Waals heterostructures. *Science (80-. ).* **360,** 1214–1218 (2018).

18. Jiang, S., Li, L., Wang, Z., Mak, K. F. & Shan, J. Controlling magnetism in 2D CrI3 by electrostatic doping. *Nat. Nanotechnol.* **13,** 549–553 (2018).

19. Zhong, D. *et al.* Van der Waals engineering of ferromagnetic semiconductor heterostructures for spin and valleytronics. *Sci. Adv.* **3,** e1603113 (2017).

20. Gong, C. *et al.* Discovery of intrinsic ferromagnetism in two-dimensional van der Waals crystals. *Nature* **546,** 265–269 (2017).

21. Huang, B. *et al.* Layer-dependent ferromagnetism in a van der Waals crystal down to the

monolayer limit. *Nature* **546,** 270–273 (2017).

22. Hara, D. J. O. *et al.* Room Temperature Intrinsic Ferromagnetism in Epitaxial Manganese Selenide Films in the Monolayer Limit. *Nano Lett.* **18,** 3125–3131 (2018).

23. Bonilla, M. *et al.* Strong roomerature ferromagnetism in VSe2monolayers on van der Waals substrates. *Nat. Nanotechnol.* **13,** 289–293 (2018).

24. Zhang, K. *et al.* Manganese Doping of Monolayer MoS2: The Substrate Is Critical. *Nano Lett.* **15,** 6586–6591 (2015).

25. Robertson, A. W. *et al.* Atomic Structure and Spectroscopy of Single Metal (Cr, V) Substitutional Dopants in Monolayer MoS2. *ACS Nano* **10,** 10227–10236 (2016).

26. Ho, C. H. *et al.* Interplay between Cr Dopants and Vacancy Clustering in the Structural and Optical Properties of WSe2. *ACS Nano* **11,** 11162–11168 (2017).

27. Kochat, V. *et al.* Re Doping in 2D Transition Metal Dichalcogenides as a New Route to Tailor Structural Phases and Induced Magnetism. *Adv. Mater.* **29,** 1703754 (2017).

28. Habib, M. *et al.* Ferromagnetism in CVT grown tungsten diselenide single crystals with nickel doping. *Nanotechnology* **29,** 115701 (2018).

29. Garcia, M. A. *et al.* Sources of experimental errors in the observation of nanoscale magnetism. *J. Appl. Phys.* **105,** 013925 (2009).

30. Geng, Y., Lee, J. H., Schlom, D. G., Freeland, J. W. & Wu, W. Magnetic inhomogeneity in a multiferroic EuTiO3 thin film. *Phys. Rev. B* **87,** 121109 (2013).

31. Duong, D. L., Yun, S. J. & Lee, Y. H. Van der Waals Layered Materials: Opportunities and Challenges. *ACS Nano* **11,** 11803–11830 (2017).

32. Zunger, A., Lany, S. & Raebiger, H. The quest for dilute ferromagnetism in semiconductors:

Guides and misguides by theory. *Physics.* **3,** 53 (2010).

33. Samarth, N. Ferromagnetic semiconductors: Battle of the bands. *Nat. Mater.* **11,** 360–361 (2012).

34. Coelho, P. M. *et al.* Room-Temperature Ferromagnetism in MoTe$_2$ by Post-Growth Incorporation of Vanadium Impurities. *Adv. Electron. Mater.* **1900044,** (2019).

35. Lachman, E. *et al.* Visualization of superparamagnetic dynamics in magnetic topological insulators. *Sci. Adv.* **1**, e1500740 (2015).

**Figure legends**

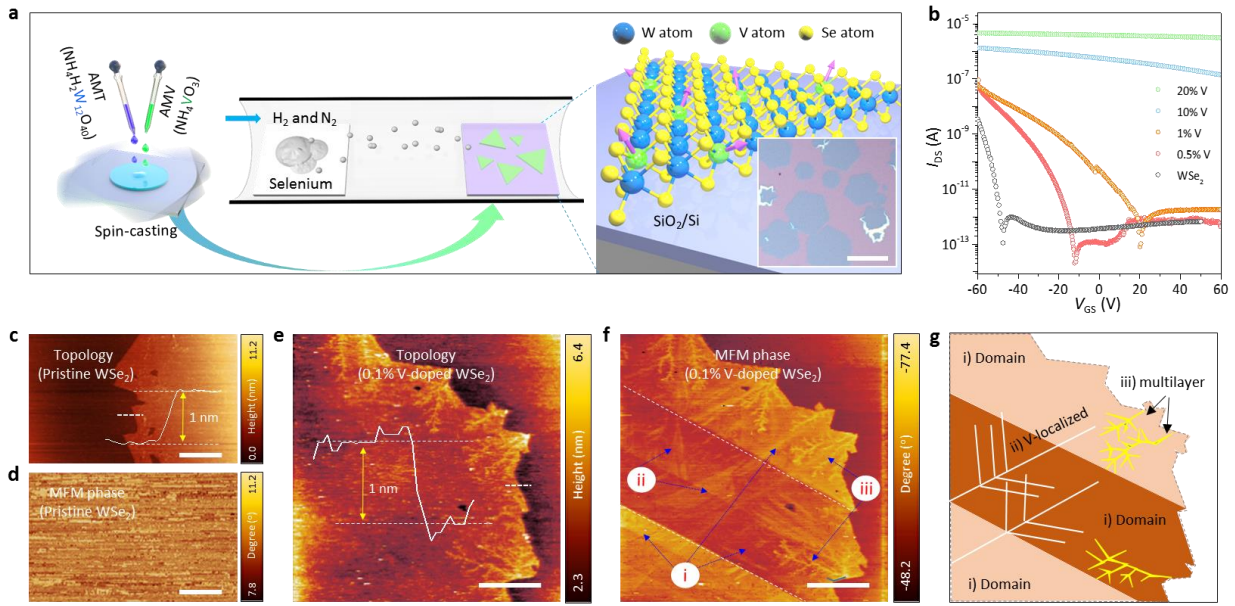

**Figure 1 | Synthesis of semiconducting V-doped monolayer WSe$_2$ and its ferromagnetic characteristics. a**, Schematic of synthesis of V-doped WSe$_2$ by mixing liquid W with V precursors. The inset shows optical image of CVD-grown V-doped WSe$_2$ monolayer. Scale bar, 10 μm **b**, Source-drain current (biased at 1V) with the gate bias for V-doped WSe$_2$ field-effect transistors with various V-doping concentrations. **c-d,** Topography (**c**) and MFM phase images (**d**) of pristine WSe$_2$ at RT. **e-f,** Topography (**e**) and MFM phase images (**f**) of 0.1% V-doped WSe$_2$ at RT. **g**, The schematic typical features observed in MFM phase images.

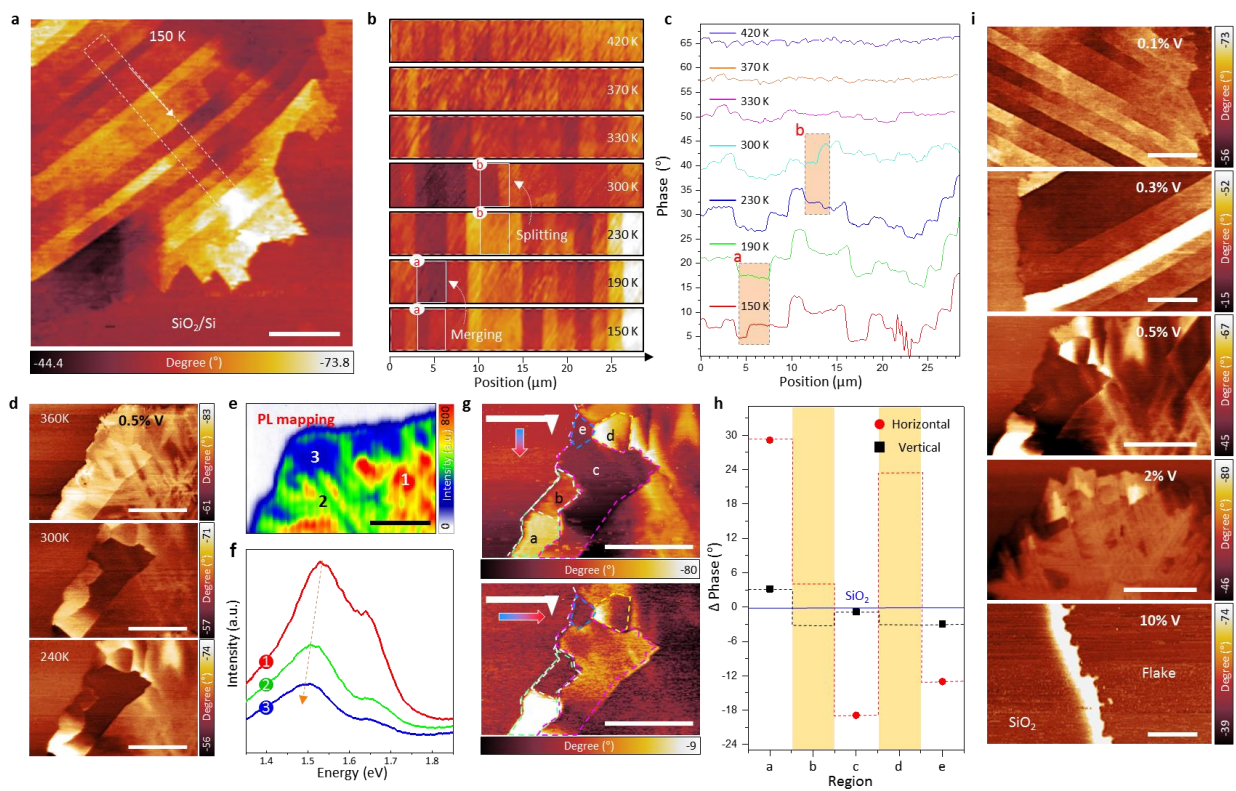

**Figure 2 | MFM with magnetic domains in 0.1% and 0.5% V-doped WSe$_2$. a,** MFM phase image of 0.1% V-doped WSe$_2$ taken at 150K. **b-c,** Temperature-dependent transition of magnetic domains (**b**) and related phase profiles (**c**) in the white-dotted box in (**a**). **d,** Temperature-dependent MFM phase images of 0.5% V-doped WSe$_2$. **e-f,** Photoluminescence mapping (**e**) and the corresponding spectra (**f**) at different positions numbered in (**e**). **g,** MFM response of 0.5% V-doped WSe$_2$ with different magnetized directions at 240K. **h,** Average phase signal of regions indicated by the letters in (**g**). The phase value for SiO$_2$ is set to zero for the reference. **i,** Magnetic domains of V-doped WSe$_2$ with different V-concentrations. All scale bars, 10 μm.

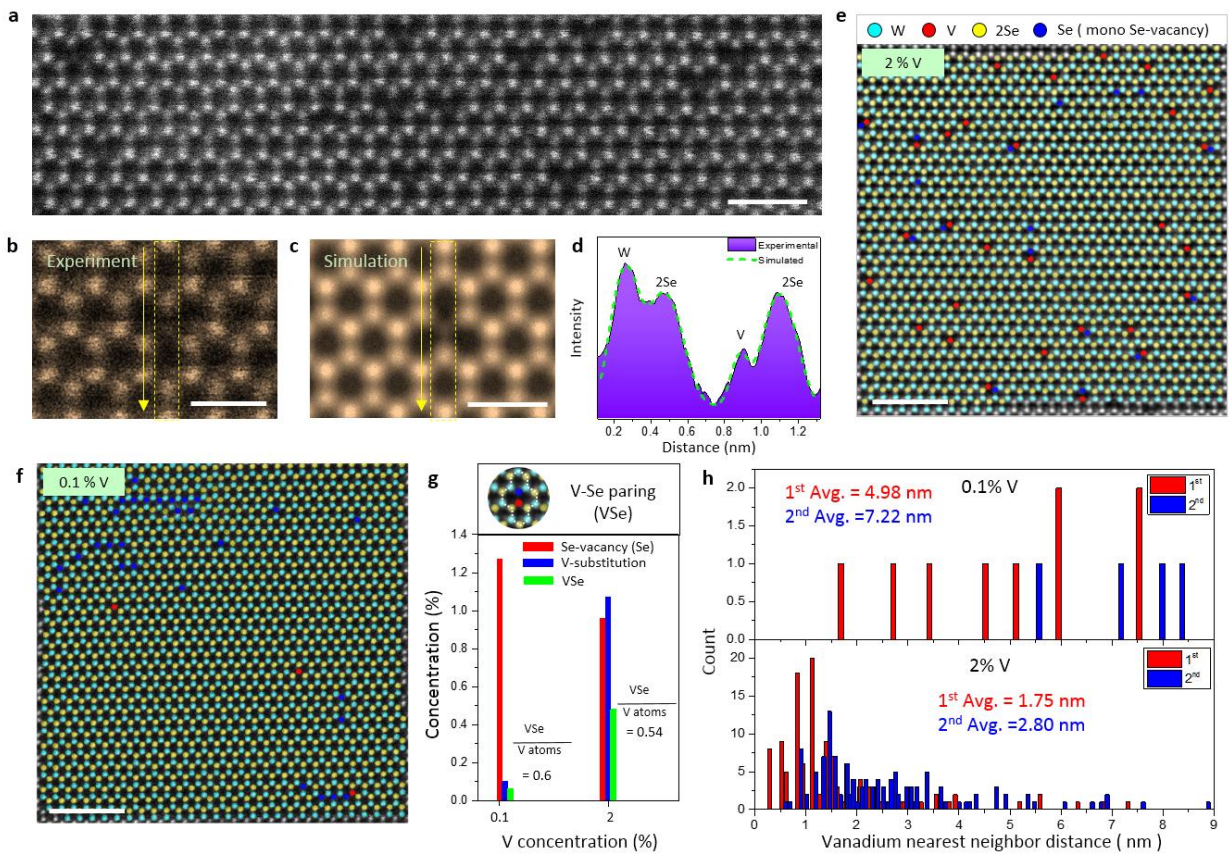

**Figure 3 | Atomic structure of V-substituted WSe₂ observed by STEM. a,** ADF-STEM images of V-doped monolayer WSe₂. Scale bar, 1 nm. **b-f,** Experimental (**b**), simulated images (**c**), and their intensity profiles (**d**) for V-doped WSe₂. Scale bar, 5 Å. **e-f,** False-color Wiener-filtered STEM images of 2% (**e**) and 0.1% (**f**) V-doped WSe₂. Scale bar, 2 nm. **g-h,** Statistical analysis of V-doped WSe₂ for V substitution, Se vacancies, VSe species (**g**) and vanadium nearest neighbor distances (**h**).

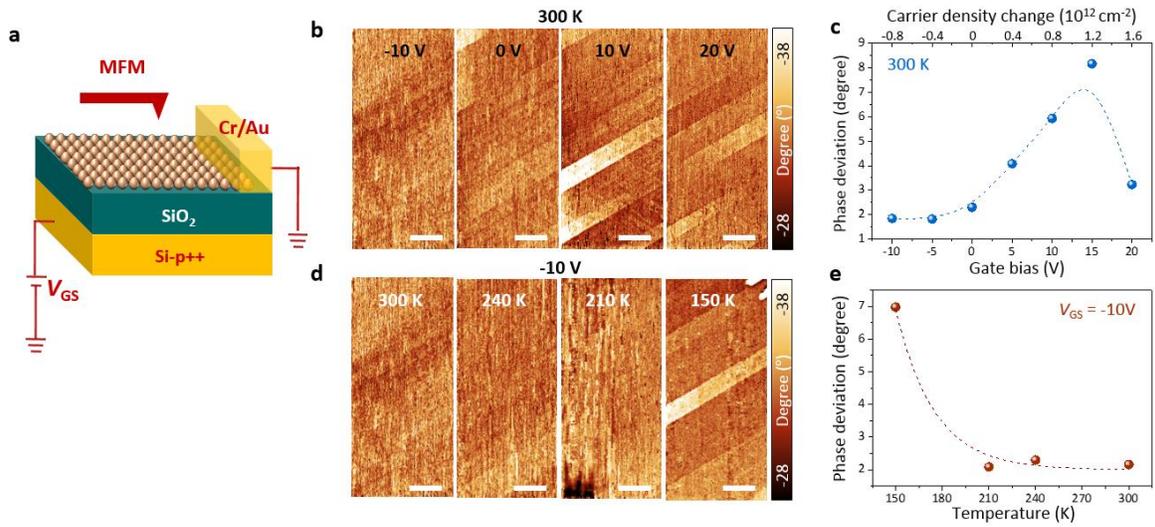

**Figure 4 | Gate-tunable magnetic properties and band structure of V-doped WSe$_2$. a,** Schematic of experimental arrangement for gate-dependent MFM measurements. **b-c,** Gate-dependent MFM images (**b**) and their phase deviation (**c**) for 0.1% V-doped WSe$_2$. **d-e,** Temperature-dependent MFM images (**d**) and their phase deviation (**e**) at -10V of gate bias. All scale bars, 10 μm.